\documentclass[twocolumn]{aastex701}
\usepackage{amsmath}
\usepackage{mathrsfs}
\usepackage{CJK}
\usepackage{multirow}
\usepackage{lipsum}

\graphicspath{{./}{figures/}}
\begin{document}
\begin{CJK*}{UTF8}{gbsn}

\title{
%Secondary Mass Function as a Key to Precision Spectral Siren Cosmology

%Sharpening Spectral Sirens: The Pairing Function Unlocks Precision Cosmology

%Mind the Secondaries: The impact of the Pairing Function on Spectral-Siren Measurements of $H_0$

Secondary-Mass Features improve Spectral-Siren $H_0$ Constraints
}

\author[0000-0001-5087-9613]{Yin-Jie Li}
\affiliation{Key Laboratory of Dark Matter and Space Astronomy, Purple Mountain Observatory, Chinese Academy of Sciences, Nanjing 210023, People's Republic of China}
\email{liyinjie@pmo.ac.cn}

\author[0000-0003-1215-6443]{Yi-Ying Wang}
\affiliation{Key Laboratory of Dark Matter and Space Astronomy, Purple Mountain Observatory, Chinese Academy of Sciences, Nanjing 210023, People's Republic of China}
\email{wangyy@pmo.ac.cn}

\author[0000-0001-9626-9319]{Yuan-Zhu Wang}
\affiliation{Institute for Theoretical Physics and Cosmology, Zhejiang University of Technology, Hangzhou, 310032, People's Republic of China}
\email{vamdrew@zjut.edu.cn}

\author[0000-0001-9120-7733]{Shao-Peng Tang}
\affiliation{Key Laboratory of Dark Matter and Space Astronomy, Purple Mountain Observatory, Chinese Academy of Sciences, Nanjing 210023, People's Republic of China}
\email{tangsp@pmo.ac.cn}

\author[0000-0002-8966-6911]{Yi-Zhong Fan}
\affiliation{Key Laboratory of Dark Matter and Space Astronomy, Purple Mountain Observatory, Chinese Academy of Sciences, Nanjing 210023, People's Republic of China}
\affiliation{School of Astronomy and Space Science, University of Science and Technology of China, Hefei, Anhui 230026, People's Republic of China}
\email[show]{The corresponding author: yzfan@pmo.ac.cn (Y.Z.F)}

\begin{abstract}

Gravitational-wave (GW) signals from compact binary coalescences (CBCs) enable independent measurements of the Hubble constant \(H_0\) via the spectral siren method, which critically depends on an accurate model of the source-frame mass distribution. While the primary mass function has been extensively studied, the impact of the secondary mass distribution on cosmological inference has been largely overlooked. Here, we perform a joint inference of population and cosmological parameters using 142 confident CBC detections from GWTC-4.0, adopting a new parametric model that flexibly describes features in both the component-mass spectrum and the pairing function, with particular emphasis on the secondary masses. We find \(H_0 = 71.4^{+13.8}_{-13.4} \;\mathrm{km\,s^{-1}\,Mpc^{-1}}\) (68\% CL) from spectral sirens alone, and \(H_0 = 73.5^{+9.2}_{-7.2} \;\mathrm{km\,s^{-1}\,Mpc^{-1}}\) when combined with the bright siren GW170817. Compared to the standard LVK Fullpop-4.0 analysis, these constraints represent improvements of \(\sim29.8\%\) and \(\sim22.2\%\) in \(H_0\) uncertainty, respectively. The enhanced precision is driven by previously unmodeled features, including peaks near \(18\,M_\odot\) and \(65\,M_\odot\) as well as mass-dependent pairing transitions at \(28\,M_\odot\) and \(52\,M_\odot\). Our results demonstrate that the secondary mass function is also a key ingredient for precision standard siren cosmology.

\end{abstract}

%\keywords{Hubble Constant, Stellar Black Hole, Gravitational Waves, LIGO/Virgo/KAGRA}

\section{Introduction}

Gravitational-wave (GW) signals from compact binary coalescences (CBCs) provide an independent and powerful method for measuring cosmological parameters, especially the Hubble constant ($H_{0}$) \citep{1986Natur.323..310S,2017Natur.551...85A,2021ApJ...908..200W,2022LRR....25....6M,2023ApJ...943...13W}. The LIGO-Virgo-KAGRA Collaboration (LVK) has relased over 200 GW events, culminating in the latest GWTC-4.0 catalog \citep{2024PhRvD.109b2001A,2023PhRvX..13d1039A,2025arXiv250818082T}. These signals offer a direct measurement of the luminosity distance ($d_{L}$) and the redshifted masses of their source components \citep{2025arXiv250904348T}, with which we can measure the cosmological expansion rate $H_{0}$, if we obtain a robust redshift estimate. The spectral siren method provides a self-contained approach by leveraging the relationship between the detector-frame and source-frame mass spectra, $m_\mathrm{det} = (1+z)m_\mathrm{src}$ \citep{2012PhRvD..85b3535T,2022PhRvL.129f1102E,2023PhRvD.108d2002M}. 

Previous population studies have focused extensively on the primary mass distribution, revealing features like a mass gap between neutron stars and black holes (BHs) \citep{2011ApJ...741..103F,2020ApJ...899L...8F,2021ApJ...923...97L,2022ApJ...931..108F}, an excess around $\sim10\,M_{\odot}$ and $\sim35\,M_{\odot}$ \citep{2021ApJ...913L..19T,2022ApJ...924..101E,2026arXiv260401420L}, mass cutoff (drop) at around $\sim 50 M_{\odot}$ \citep{2021ApJ...913...42W}, as well as peaks near $\sim 65M_\odot$ \citep{2024MNRAS.527..298T,2025arXiv250819208M}, possibly near $18 M_\odot$ \citep{2023ApJ...955..107F,2023ApJ...946...16E,2024ApJ...960...65S,2024PhRvX..14b1005C,2025arXiv251025579T,2026arXiv260401420L}.
Additionally, the features in the secondary-mass function (or mass-ratio distribution or pairing function) are also revealed. For example, the mass-ratio distributions are different in the mass range between $m_1\sim29M_\odot$ and $m_1\sim 50 M_\odot$\citep{2022ApJ...933L..14L,2025arXiv250915646B}, orientated from different BBH subpopulations\citep{2024ApJ...977...67L,2026arXiv260317987R}.
Recent work have also reveled signatures of mass cutoff/drop in the secondary-mass function \citep{2025arXiv250904151T,2026ApJ...998L..20R,2026arXiv260505563X}. 

Features in the component-mass function, especially in the primary-mass function, have already carefully modeled for spectral sirens \citep{2019ApJ...883L..42F,2021ApJ...908..215Y,2025arXiv250904348T,2025ApJ...978..153F,2026arXiv260103347T,2025arXiv250903607M,2026arXiv260103257P}.
Additional features in the mass functions of subpopulations are also revealed  \citep{2022ApJ...941L..39W,2024PhRvL.133e1401L,2025ApJ...987...65L,2025arXiv250923897L,2025PhRvL.134a1401A,2025arXiv250904637A,2025arXiv250909123A,2025arXiv251022698W,2025arXiv251105316T,2026arXiv260107908P}, which are also helpful to improve the measurement of $H_0$, so-called, Multi-spectral sirens \citep{2024arXiv240606109U,2024ApJ...976..153L,2025ApJ...985..220T}.

However, the secondary mass function, has not modeled carefully for spectral sirens yet. Given that the spectral siren method is benefit from features of the entire CBC mass distribution \citep{2022PhRvL.129f1102E,2021PhRvD.104f2009M,2024PhRvD.109h3504P}, neglecting structure in the secondary mass function could discard valuable cosmological information.

In this Letter, we analyze the GWTC-4.0 catalog using 142 confident CBC detections and new mass models that allow for additional features in the component-mass spectrum and pairing function. In Section~\ref{sec:models}, we introduce the cosmological model and population models. In Section~\ref{sec:result}, we presents the $H_0$ measured with our population models, and demonstrate how our models improve the measurement of Hubble constant relative to the Fullpop-4.0 and \textsc{Multi-Peaks} (MLTP) of LVK's analysis \citep{2025arXiv250904348T}.  Finally, we carry out discussion and make our conclusion in Section~\ref{sec:conclusion}.

\section{Method}\label{sec:models}
Following \citet{2025arXiv250904348T}, we employ a general and well-established method (i.e., Hierarchical Bayesian Inference, see Appendix~\ref{app:method}) to measure the Hubble constant as well as the mass distribution of GW events.

\subsection{Cosmological model}
We adopt a flat $\Lambda$CDM cosmological model and assume a constant dark energy density throughout cosmic expansion. The luminosity distance $D_L$ as a function of redshift $z$ is then given by \citep{2023ApJ...949...76A,2025arXiv250904348T}
\begin{equation}
\begin{aligned}
D_L(z)&=\frac{c(1+z)}{H_0}\int_{0}^{z}{[\Omega_{\rm m}(1+x)^3+1-\Omega_{\rm m}]^{-1/2} \mathrm{d}x}\\
&={\rm F}(z|H_0,\Omega_{\rm m}),
\end{aligned}
\end{equation}
where $\Omega_{\rm m}$ denotes the present-day dimensionless matter density, and $H_0$ is the Hubble constant.
Gravitational-wave signals allow measurements of the detector-frame masses of BBHs and the luminosity distance (i.e., $M_1$, $M_2$, $D_{L}$). Given the cosmological parameters $H_0$ and $\Omega_{\rm m}$, the source-frame masses can be obtained via the relation $m_{1,2} = M_{1,2} / (1 + z(D_L)) = M_{1,2} / (1 + {\rm F}^{-1}(D_L|H_0, \Omega_{\rm m}))$.
Following \citet{2025arXiv250904348T}, for all the analysis we fix $\Omega_{\rm m}=0.3065$.

\subsection{Population models for CBCs}
To analyze with 142 CBCs, we construct a population model (named Flexible-CBC) building on the FullPop-4.0 model of LVKC \citep{2025arXiv250904348T}. Our population model has a semi-parametric formula, and the component-mass distribution (unpaired) is described by
\begin{equation}
\begin{aligned}
p_{\rm cbc}(m|\Lambda) \propto &\mathcal{S}(m|\Lambda_s) e^{f(m|\{n_i\}_{i=1}^{15})} \\  &\times \mathcal{BPD}(m | m_{\rm min},m_{\rm max},\alpha_1,\alpha_2,m_{\rm b})
\end{aligned}
\end{equation}
where $\mathcal{BPD}$ and $\mathcal{S}$ are the \textsc{PowerLawDipBreak} model and notch filters as defined by Eq.~(C32-C36) of \citet{2025arXiv250904348T} originating from \citet{2020ApJ...899L...8F} and \citet{2022ApJ...931..108F}. $f(m|\{n_i\}_{i=1}^{15})$ is the cubic spline function to describe the underlying features beyond \textsc{PowerLaw} as first used by \cite{2022ApJ...924..101E}.
Then the Flexible-CBC mass function is 
\begin{equation}
\begin{aligned}
p_{\rm CBC}(m_1,m_2|&\Lambda) \propto  p_{\rm cbc}(m_1|\Lambda) p_{\rm cbc}(m_2|\Lambda)\\
 &\times f_{\rm cbc}(m_1,m_2|\Lambda)
\end{aligned}
\end{equation}
with
\begin{equation}\label{eq:md}
f_{\rm cbc}(m_1,m_2|\Lambda)=
\begin{cases}
(\frac{m_2}{m_1})^{\beta_3},~~~~m_{\rm d,l}<m_1<m_{\rm d,h}, \\
f(m_1,m_2| \beta_1,\beta_2,m_{\rm break}),~{\rm else.}
\end{cases}
\end{equation}
where $f(m_1,m_2| \beta_1,\beta_2,m_{\rm break})$ is the pairing function as defined by Eq.~C37 in \citet{2025arXiv250904348T},
and we introduce another pairing index for the events with primary masses within $[m_{\rm d,l}, m_{\rm d,h}]$, such a configuration is motivated by the previous work for the mass-dependent pairing function (or correlation between primary masses and mass ratio) \citep{2022ApJ...933L..14L,2025arXiv250915646B,2024ApJ...977...67L,2026arXiv260317987R}.

The Fullpop-PS model is reduced from Flexible-CBC, where we set $\beta_3\equiv\beta_2$, i.e., we use a constant pairing function for all the BBHs.

\subsection{Population models for BBHs}\label{app:BBH_model}
The component mass function for the BBHs is simpler than that for CBC, which employs a \textsc{PowerLawSpline} model as used in \citet{2022ApJ...924..101E,2023PhRvX..13a1048A},
\begin{equation}
p_{\rm bbh}(m|\Lambda) \propto  \mathcal{SP}(m | m_{\rm min},m_{\rm max},\alpha,\delta_{\rm m})e^{f(m|\{n_i\}_{i=1}^{15})}
\end{equation}
Then the Flexible-BBH mass function reads,
\begin{equation}
\begin{aligned}
p_{\rm BBH}(m_1,m_2|&\Lambda) \propto  p_{\rm bbh}(m_1|\Lambda) p_{\rm bbh}(m_2|\Lambda)\\
 &f_{\rm bbh}(m_1,m_2|\Lambda)
\end{aligned}
\end{equation}
with
\begin{equation}
\begin{aligned}
f_{\rm bbh}&(m_1,m_2|\Lambda)=
\begin{cases}
(\frac{m_2}{m_1})^{\beta_3}  & , m_{\rm d,l}<m_1<m_{\rm d,h}, \\
(\frac{m_2}{m_1})^{\beta_2}  & , {\rm else.}
\end{cases}
\end{aligned}
\end{equation}

We also investigate a PS paired model, which is reduced from the Flexible-BBH when set $\beta_3\equiv\beta_2$. Besides, the \textsc{MultiPeak} (MLTP) model of \citet{2025arXiv250904348T} and its pairing version, i.e., the MLTP paired model, are also analysed for comparison; the latter is expressed as
\begin{equation}
\begin{aligned}
p_{\rm MP-paired}(m_1,m_2|&\Lambda) \propto  p_{\rm mltp}(m_1|\Lambda) p_{\rm mltp}(m_2|\Lambda)(\frac{m_2}{m_1})^{\beta},
\end{aligned}
\end{equation}
where $p_{\rm mltp}(m_1|\Lambda)$ is defined by the Eq.~C29 in \citet{2025arXiv250904348T}.

\section{Results\label{sec:result}}

In this section, we report our best $H_0$ measurement derived from 142 CBC events using the Flexible-CBC (fiducial) model. This model incorporates flexible prescriptions for the underlying structures of the component-mass function and the pairing function. When presenting our results, we quote median values along with $68.3\%$ ($90\%$) symmetric credible intervals. We then compare the fiducial model with other simpler models, as well as with BBH models based on 137 BBH events, to illustrate the impact of the additional features captured by the fiducial model on the $H_0$ measurement.
Table~\ref{tab:reweighed_BF} present the Bayes factors of population models relative to the Fullpop-4.0 model or MLTP model.

\begin{table}[htpb]
\centering
\caption{Bayes factors of population models relative to the Fullpop-4.0 model or MLTP model}\label{tab:reweighed_BF}
\begin{tabular}{cc}
\hline
\hline
 Population mdoel & $\ln\mathcal{B}$ \\
\hline
\multicolumn{2}{c}{\bf CBC mass function (142 events)} \\
Fullpop-4.0 & 0 \\
Fullpop-PS & 5.7\\
Flexible-CBC (fiducial) & 7.4 \\
\hline
\multicolumn{2}{c}{\bf BBH mass function (137 events)} \\
MLTP & 0 \\
MLTP paired & 6.7\\
PS paired & 10.4 \\
Flexible-BBH & 11.5 \\
\hline\hline
\end{tabular}
\end{table}

\subsection{The Hubble constant\label{sec:hubbleresult}}
\begin{figure*}[ht!]
\centering
  \includegraphics[width=0.48\textwidth]{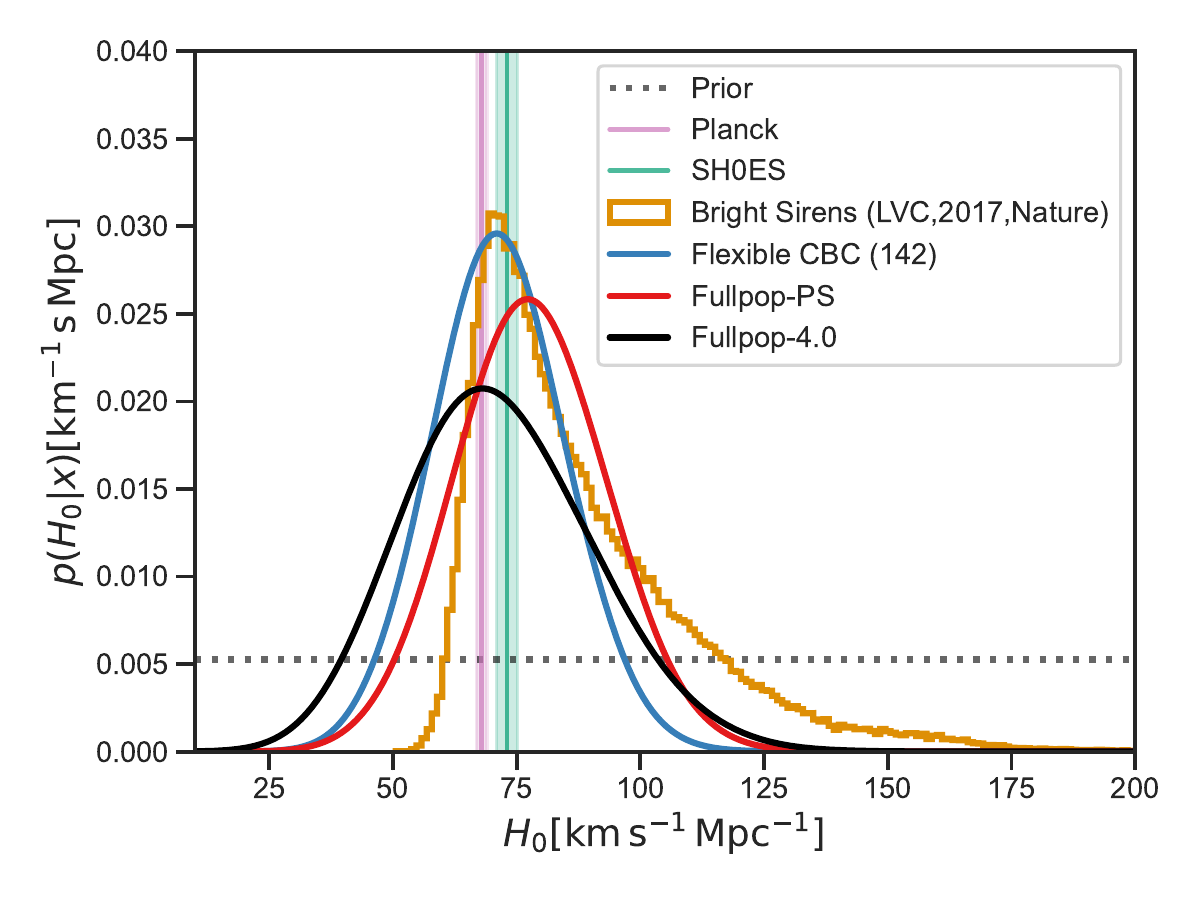}
  \includegraphics[width=0.48\textwidth]{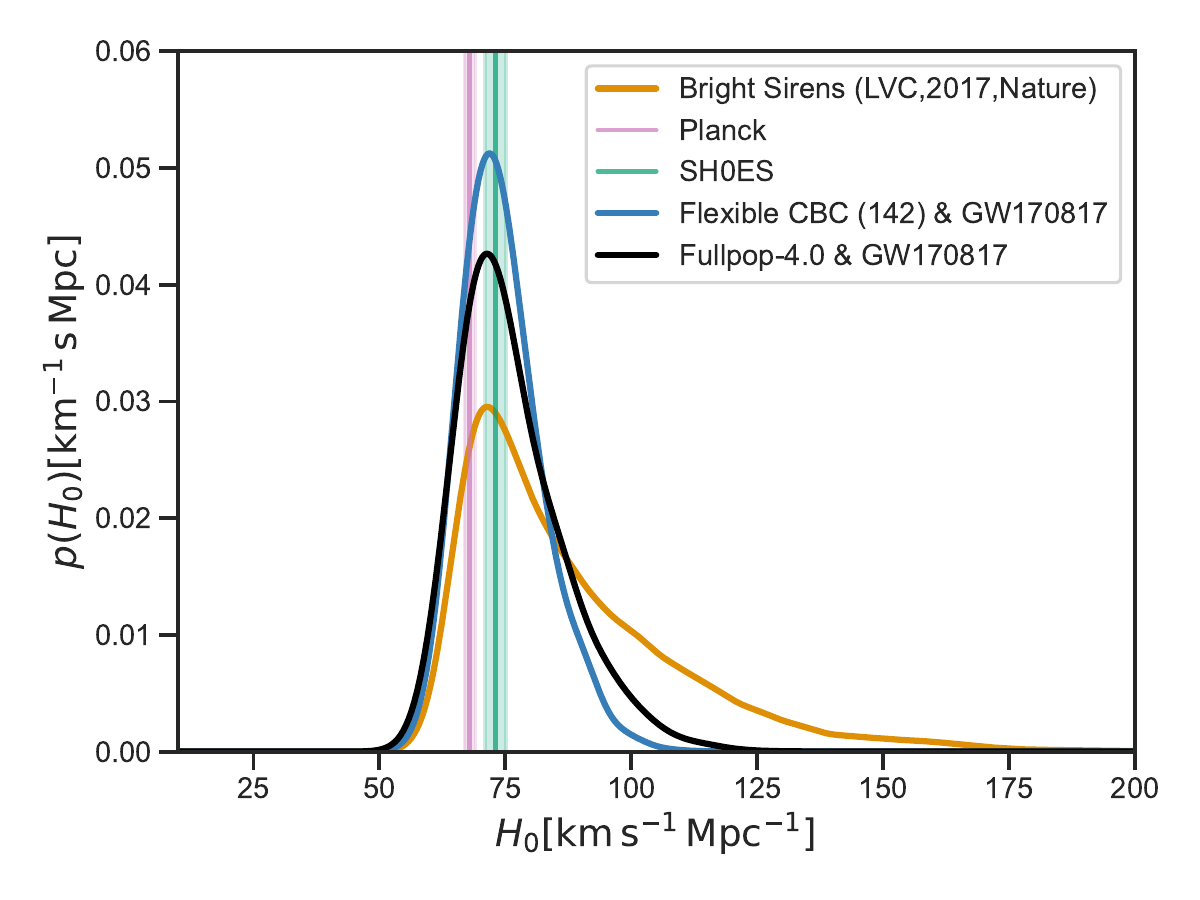}
  \caption{Hubble constant posteriors. Left: Hubble constants inferred with Flexible-CBC, Fullpop-PS, Fullpop-4.0 via spectral sirens of 142 events, as well as that inferred via bright siren of GW170817 \citep{2017Natur.551...85A}. Right: Results of Flexible-CBC and Fullpop-4.0 combined with bright siren of GW170817, comparing to the result of bright siren solely. Vertical lines indicate the Hubble tension reference values from Planck and SH0ES \citep{2020A&A...641A...6P,2022ApJ...934L...7R}.}
     \label{Fig:H0_posterior}
\end{figure*}

Figure~\ref{Fig:H0_posterior} shows the posteriors of the Hubble constant obtained with 142 CBCs using our new mass models, compared to the LVK's Fullpop-4.0 model. Our Flexible-CBC model yields a posterior of $H_{0}=71.4^{+13.8}_{-13.4}\,(71.4^{+22.8}_{-21.9})\,\text{km\,s}^{-1}\,\text{Mpc}^{-1}$. This corresponds to a {$29.8\%$} improvement in the Hubble constant constraint compared to the Fullpop-4.0 model. This improvement attributes to the reconstruction of the additional features in the mass spectrum and pairing function, which we discuss in Section~\ref{sec:massscales}. We Note that the uncertainty of the $H_0$ inferred with our model via spectral siren is comparable to that inferred with bright siren from GW170817 \citep{2017Natur.551...85A}.
Combining the constraints of Flexible-CBC model with the bright siren GW170817 yields $H_{0}=73.5^{+9.2}_{-7.2}\,\text{km\,s}^{-1}\,\text{Mpc}^{-1}$ showing also improvement of {22.2\%} over the combination of Fullpop-4.0 with bright siren, i.e., the best constraint in analysis of \citet{2025arXiv250904348T}.

\subsection{Features beyond Fullpop-4.0}\label{sec:massscales}
\begin{figure*}[ht!]
\centering
  \includegraphics[width=0.9\textwidth]{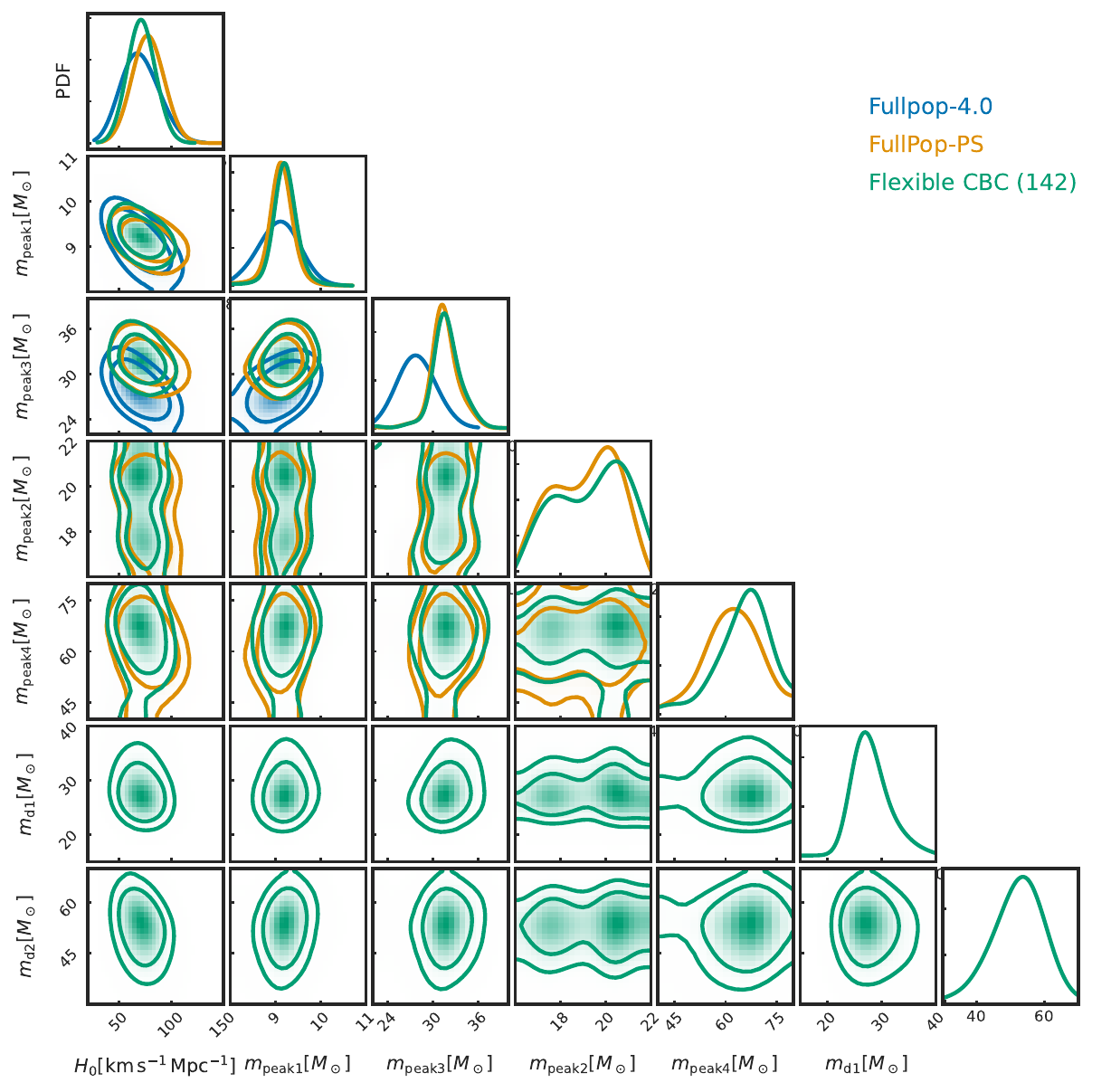}
  \caption{Posterior distributions of the Hubble constant as well as parameters that describe the CBC mass distribution, for Flexible-CBC, Fullpop-PS, and Fullpop-4.0 models. The four peaks are the local maximum points of the cubic spline function for the Flexible-CBC and Fullpop-PS. Note that the peak1 and the peak3 are for the center values of the two Gaussian ($\mu_{\rm g}^{\rm low}$ and $\mu_{\rm g}^{\rm high}$) defined in the Fullpop-4.0 \citep{2025arXiv250904348T}.}
     \label{Fig:CBC_corner}
\end{figure*}

The improvement in the $H_0$ measurement result from the modeling of the previously un-modeled features in the pairing function and the component-mass function, i.e., the peaks at $\sim 65 M_{\odot}$ and (possibly) at $\sim 18M_\odot$, as well as the mass dependent pairing function that transitions at $\sim 28 M_\odot$ and $\sim 53 M_\odot$, which are correlated with $H_0$ (see Figure~\ref{Fig:CBC_corner}). These features are not present when using the standard \textsc{FullPop-4.0} model\citep{2025arXiv250904348T}.
The additional mass peaks, are consistent with that previously reported in some other population analyses \citep{2023ApJ...955..107F,2023ApJ...946...16E,2024ApJ...960...65S,2024PhRvX..14b1005C,2025arXiv251025579T,2026arXiv260401420L}, and the features in the mass-dependent pairing function also agree with those found in the studies on mass-ratio distribution \citep{2022ApJ...933L..14L,2025arXiv250915646B}.

\begin{figure}[ht!]
\centering
  \includegraphics[width=0.5\textwidth]{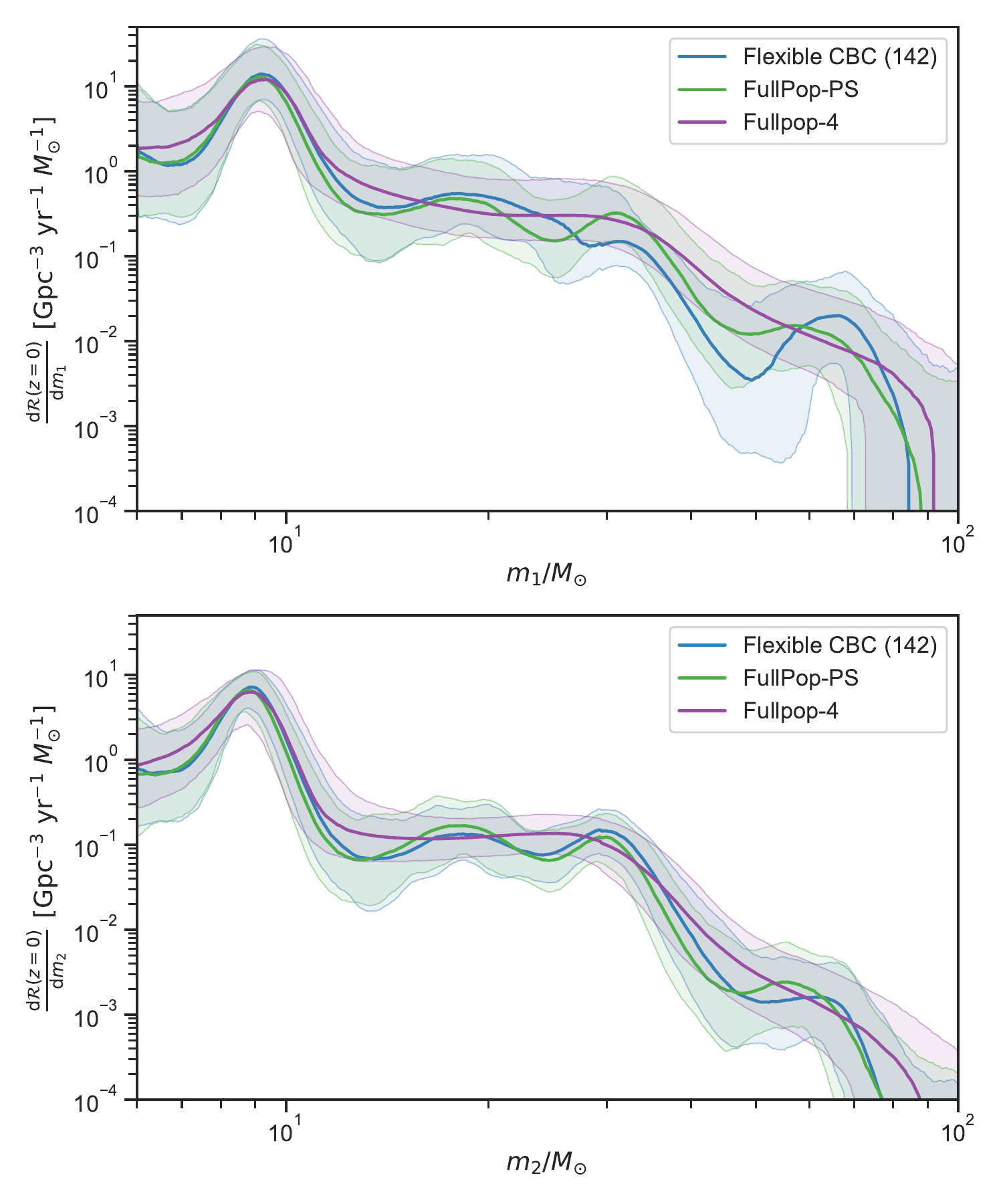}
  \caption{Mass distribution of primary and secondary BHs. The solid lines and shaded areas are for median values and 90\% credible intervals.}
     \label{Fig:Fullpop_mass}
\end{figure}

We plot the mass-dependent pairing parameters and the local maximum points in the cubic spline functions (for the Flexible-CBC and Fullpop-PS), together with the peak parameters ($\mu_{\rm g}^{\rm low}$ and $\mu_{\rm g}^{\rm high}$) of Fullpop-4.0, in Fig.~\ref{Fig:CBC_corner}, which illustrate the impact of these features on the $H_0$ measurement. Besides the peaks at $\sim 10M_\odot$ and $\sim 32M_\odot$ that were already modeled by Fullpop-4.0~\citep{2025arXiv250904348T}, additional peaks at $\sim 65 M_{\odot}$ and (possibly) at $\sim 18M_\odot$ are also correlated with $H_{0}$. This highlights the advantage of Flexible-CBC and Fullpop-PS over Fullpop-4.0, owing to their flexibility in the mass function. 
On the other hand, transition points for the mass-dependent pairing correlate with $H_0$ as well. This highlights the advantage of the Flexible-CBC model over Fullpop-PS, since it better captures the pairing mechanism of CBCs and thereby improves the modeling of the primary and secondary mass functions. We therefore emphasize that careful modeling of either the secondary mass function or the mass-dependent pairing function is essential for achieving higher precision in constraining $H_0$ with spectral sirens.

Figure~\ref{Fig:Fullpop_mass} shows the primary and secondary mass distributions inferred with Flexible-CBC, Fullpop-PS, and Fullpop-4.0. As illustrated above, the Flexible-CBC and Fullpop-PS models exhibit more peaks in the component mass functions than Fullpop-4.0. We notice that Flexible-CBC reveals different behaviors between its primary and secondary mass functions  --  a feature not observed in Fullpop-PS. Table~\ref{tab:reweighed_BF} presents the Bayes factors comparing these models. Both Flexible-CBC and Fullpop-PS are more favored than Fullpop-4.0, with Bayes factors of $\ln\mathcal{B}=5.7$ and $\ln\mathcal{B}=7.4$, respectively. We therefore attribute the improvement in the $H_0$ measurement achieved by our Flexible-CBC model over the standard LVK Fullpop-4.0 to its better modeling of the CBC mass functions.

\subsection{The impact of the pairing function}

The above results indicate that the improvement in $H_0$ can arise not only from the component-mass function but also from the pairing function. For further illustration, we carry out an analysis using only 137 BBHs (i.e., excluding five potential NS events), since the improvement in our model stems solely from better modeling of the BBH mass function. The dedicated BBH mass functions  --  named Flexible-BBH, PS paired, and MLTP paired  --  are built upon the \textsc{MltiPeaks} (MLTP) model of \citet{2025arXiv250904348T} and are defined in Appendix~\ref{app:BBH_model}. These models are inferred from 137 BBHs with FAR $<0.25/{\rm yr}$ from GWTC-4.

MLTP paired extends the MLTP model by incorporating pairing. PS paired adopts a more flexible component-mass function than MLTP paired, and Flexible-BBH further introduces a mass-dependent pairing function relative to PS paired. With successively increasing capabilities, these models demonstrate how each of our improvements enhances the precision of $H_0$ measurement. Figure~\ref{Fig:BBH_H0}, Figure~\ref{Fig:BBH_mass}, and Figure~\ref{Fig:BBH_corner} show, respectively, the marginalized distribution of $H_0$, the reconstructed mass functions, and the posterior distributions of model parameters. We find that the Flexible-BBH and PS paired models are more favored than MLTP paired, with Bayes factors of $\ln\mathcal{B}=4.8$ and $\ln\mathcal{B}=3.7$, respectively.

Interestingly, MLTP paired is more favored than MLTP by a Bayes factor of $\ln\mathcal{B}=6.7$, and the $H_0$ measured with MLTP paired shows a 30.2\% improvement relative to that from MLTP. This result highlights the importance of properly modeling the secondary mass distribution, as there are features in the $m_2$ mass distribution that cannot be described simply by a power-law distribution conditioned on primary masses. Figure~\ref{Fig:BBH_mass} compares the inferred secondary mass distributions across different population models, showing that MLTP struggles to capture the underlying structures that other models recover  --  for example, the peaks at $\sim 9M_\odot$ and $\sim 32M_\odot$. We have also carried out a posterior predictive check, presented in Figure~\ref{Fig:PPC}, which shows that the observed band from MLTP lies slightly outside the predicted region, indicating the weakness of MLTP compared to other models. In contrast, MLTP paired, PS paired, and Flexible-BBH perform progressively better.

We note that in the LVK analysis, the Hubble constant measured with Fullpop-4.0 (142 CBC events) is significantly better than that with MLTP (137 BBH events), yielding an improvement of $>50\%$ in $H_0$, despite both models having two peaks in the BBH mass function. Based on our illustration above, this improvement arises not only from the additional five low-mass events but also from the pairing function (i.e., the secondary mass function) adopted by Fullpop-4.0. Specifically, the improvement of Fullpop-4.0 over MLTP paired is 21.5\%, attributed to the five extra low-mass events, while the $H_0$ measurement from Flexible-CBC shows only a 7.5\% improvement over Flexible-BBH (see Figure~\ref{Fig:CBC_BBH_H0} in Appendix~\ref{app:results}).

\section{Discussion and Conclusions \label{sec:conclusion}}
In this work, we have shown that additional features in the component-mass function and the pairing function (and hence the secondary-mass function) play a crucial and previously under-appreciated role in spectral-siren cosmology. By employing a population model that relies on a flexible component-mass function and a mass-dependent pairing function, we have identified new peaks in the component-mass distribution at $\sim18\,M_{\odot}$ and $\sim65\,M_\odot$, as well as transitions in the pairing function at $m_1\sim 27.6^{+7.0}_{-3.6}\,M_\odot$ and $\sim52.9^{+9.7}_{-12.6}\,M_\odot$. These features significantly improve the constraining power on the Hubble constant, yielding a \textbf{$29.8\%$} improvement with 142 CBC events relative to the latest LVK results (Fullpop-4.0) from GWTC-4.0.

We have also demonstrated the importance of modeling the pairing function (and hence the secondary-mass function) for BBH population analysis and spectral sirens. Modeling the secondary-mass distribution simply as $\propto m_2^\beta$ conditioned on $m_1$ risks masking or mis-specifying features in $m_2$, thereby biasing the $H_0$ measurement. A more flexible pairing function  --  for example, one that varies with primary mass  --  enhances the flexibility of the $m_1$-$m_2$ distribution, revealing additional features in the mass function and improving the $H_0$ measurement. Future standard-siren analyses should therefore adopt more flexible models capable of capturing structures in both the component-mass distribution and the pairing function. As the number of gravitational-wave detections continues to grow, the secondary-mass spectrum will provide increasingly rich features for precision cosmology, helping to shed light on the Hubble tension and the expansion history of the Universe.
	 	
The peaks, gaps/dips, and cutoffs/drops in the component mass, primary mass, or chirp mass distributions of BBHs have been investigated using various approaches \citep[e.g.][]{2021CQGra..38o5007T,2021ApJ...917...33L,2022ApJ...931..108F,2022MNRAS.509.5454R,2022ApJ...924..101E,2022PhRvD.105l3014S,2023ApJ...957...37R,2025A&A...694A.186G,2025PhRvD.111f3043H,2024PhRvX..14b1005C}. The peaks identified in this work have been previously reported in several studies, and some of them have been introduced to spectral-siren cosmology \citep[e.g.][]{2025arXiv250904348T,2025ApJ...978..153F,2026arXiv260103347T,2025arXiv250903607M,2026arXiv260103257P,2026arXiv260414290G}. These peaks likely originate from different formation channels: for instance, the $\sim 9\,M_\odot$, $\sim 32\,M_\odot$, and $\sim 65\,M_\odot$ peaks may be associated with isolated channels, dynamically first-generation channels, and second-generation mergers, respectively \citep{2022ApJ...941L..39W,2023arXiv230401288G,2024PhRvL.133e1401L,2024ApJ...977...67L,2024A&A...692A..80P,2025PhRvL.134a1401A,2026arXiv260317987R,2026arXiv260320430A,2026arXiv260407456G}; see also alternative interpretations \citep{2025arXiv250915646B,2025CQGra..42v5008K}.

Nevertheless, features in the pairing function (or secondary-mass function) are introduced and emphasized for spectral sirens -- for the first time. The presence of such features -- specifically, a distinct mass-ratio distribution for $m_1\sim 28-53\,M_\odot$ -- can be important for determining $H_0$. This feature may be interpreted as a signature of a specific formation channel, such as dynamically formed first-generation BBHs in star clusters~\cite{2016PhRvD..93h4029R,2019MNRAS.486.5008A}, or a sub-population formed through isolated binary evolution with different initial conditions (e.g., chemically homogeneous evolution~\cite{2016MNRAS.460.3545D,2016A&A...588A..50M}). Beyond these features, there may also be correlations between other parameters and the mass of CBCs~\citep{2021ApJ...922L...5C,2022ApJ...932L..19B,2022ApJ...928..155T,2024PhRvD.109j3006H,2024arXiv240601679P,2024ApJ...977...67L,2025ApJ...987...65L,2024arXiv241019145C,2026ApJ...996...71H,2025arXiv250620731A,2025PhRvD.111f1305H,2026A&A...708A..62W,2026PhRvD.113d3048B,2026arXiv260211030Z,2026ApJ...999L..30V}, which may also contribute to spectral-siren cosmology~\citep[see, e.g.,][]{2024ApJ...976..153L,2024arXiv240606109U,2025ApJ...985..220T}.

Furthermore, the redshift evolution of the merger rate may also depend on BBH masses~\citep{2024ApJ...975...54G,2025PhRvD.111l3046G,2025A&A...702A..52R,2026arXiv260103456F}. As the sensitivity of gravitational-wave detectors continues to improve, more flexible population models should be constructed to fully exploit the available information, revealing finer features in the GW event population and enabling more precise measurements of the expansion of the universe~\citep{2022PhRvR...4a3247Z,2023PhRvD.108d2002M,2025ApJ...986...61L,2024arXiv240517161P}. Additionally, the stochastic gravitational-wave background offers a complementary route for measuring the Hubble constant when combined with resolved GW events, an approach known as stochastic sirens~\citep{2025A&A...701A..36F,2026PhRvL.136j1003C}, since its amplitude is sensitive to the redshift evolution of the CBC population~\citep{2020ApJ...896L..32C}. Gravitational waves are thus opening a new window into precision cosmology, delivering ever-increasing precision and helping to address critical issues such as dynamical dark energy~\citep{2006IJMPD..15.1753C,2013FrPhy...8..828L,2025arXiv251111795P,2026A&A...707A.189W} and the Hubble tension~\citep{2024hct..book.....D,2025PDU....4901965D,2022PhRvD.106b3011W,2022PhRvL.129f1102E}.

\begin{acknowledgments}
Yin-Jie Li thanks Gr\'egoire Pierra for the helpful discussions about reproducing the LVKC's results \citep{2025arXiv250904348T}. This work is supported in part by the National Natural Science Foundation of China (No. 12233011, No. 12588101, No. 12503059, No. 12203101, No. 12303056),  the New Cornerstone Science Foundation through the EXPLORER PRIZE, the General Fund (No. 2024M753495, No. 2025M783236) and the Postdoctoral Fellowship Program (No. GZB20250738) of the China Postdoctoral Science Foundation, and the Priority Research Program of the Chinese Academy of Sciences (No. XDB0550400). This research has made use of data and software obtained from the Gravitational Wave Open Science Center (https://www.gw-openscience.org), a service of LIGO Laboratory, the LIGO Scientific Collaboration and the Virgo Collaboration. LIGO is funded by the U.S. National Science Foundation. Virgo is funded by the French Centre National de Recherche Scientifique (CNRS), the Italian Istituto Nazionale della Fisica Nucleare (INFN) and the Dutch Nikhef, with contributions by Polish and Hungarian institutes. 
\end{acknowledgments}

\software{Bilby \citep[version 1.1.4, ascl:1901.011,  \url{https://git.ligo.org/lscsoft/bilby/}][]{2019ascl.soft01011A},
          %Dynesty \citep[version 1.0.1, \url{https://github.com/joshspeagle/dynesty}]{2020MNRAS.493.3132S},
          PyMultiNest \citep[version 2.11, ascl:1606.005, \url{https://github.com/JohannesBuchner/PyMultiNest}][]{2016ascl.soft06005B},
          %Precession \citep[version 1.0.3, \url{https://github.com/dgerosa/precession} ]{2016PhRvD..93l4066G}
          }

\appendix

\twocolumngrid

\section{Hierarchical Bayesian Inference}\label{app:method}
Given a population distribution $\Lambda$, the likelihood of the GW data $\{d\}$ from $N_{\rm det}$ detections is \cite{2025arXiv250904348T}, 
\begin{equation}
\mathcal{L}(\{d\}|\Lambda)\propto N^{N_{\rm det}}e^{-N_{\rm exp}} \prod_{i}^{N_{\rm det}}{\int{\pi(\theta_i|\Lambda)\mathcal{L}(d_i|\theta_i)d\theta_i}},
\end{equation}
where $N=\int{R(z|\Lambda)\frac{dV_{\rm c}}{dz}\frac{T_{\rm obs}}{1+z}dz}$ is the total number of binary black hole mergers in the surveyed volume, and $N_{\rm exp}=N\int{P({\rm det}|\theta)\pi(\theta|\Lambda)d\theta}$ is the expected number of detections, with detection probability $P({\rm det}|\theta)$. $N_{\rm exp}$ is evaluated using a Monte Carlo integral over the public injection set \cite{2025arXiv250818083T, 2025PhRvD.112j2001E} (Adopted from \href{https://zenodo.org/records/16740128}{Zenodo}), and $\mathcal{L}(d_i|\theta_i)$ is computed from the posterior samples Following \cite{2025arXiv250904348T}, we enforce a total Monte Carlo uncertainty $\sigma_{\rm tot} < 1$ to ensure reliable likelihood estimation. We employ the \texttt{Pymultinest} sampler \cite{2016ascl.soft06005B} to sample the hyperparameter posterior distribution.
The priors of the parameters of all the models are summarized in Table.~\ref{tab:prior}.

\begin{table*}[htpb]
\centering
\caption{Summary of model parameters.}\label{tab:prior}
\begin{tabular}{lcccc}
\hline
\hline
Parameter     &  Description & Prior \\
\hline
$m_{\rm min}[M_{\odot}]$ / $m_{\rm max}[M_{\odot}]$   & The minimum / maximum  mass & $U(0.4,1.4)$ / $U(50,200)$  \\
$\alpha_1$ / $\alpha_2$ & Slope index of the power-law before / after $b$ & $U(-4,12)$ \\
$\beta_1$ / $\beta_2$ & Spectral index of the pairing function before / after $m_{\rm break}$ & $U(-4,12)$ \\
$m_{\rm d,l}[M_{\odot}]$ / $m_{\rm d,h}[M_{\odot}]$ & Defined in Eq.~(\ref{eq:md}) & $U(15,40)$ / $U(30,70)$ \\
$\beta_3$ & Spectral index of the pairing function for $m_{\rm d,l}<m_1<m_{\rm d,h}$ & $U(-4,12)$ \\
$\delta_{\rm m}^{\rm min}[M_{\odot}]$ / $\delta_{\rm m}^{\rm max}[M_{\odot}]$ & Smooth scale of the mass lower / upper edge & $LU(10^{-2},1)$  \\
$m_{\rm d}^{\rm min}[M_{\odot}]$ / $m_{\rm d}^{\rm max}[M_{\odot}]$ & Lower / Upper edge of the dip& $U(1.5,3)$ / $U(5,9)$  \\
$\delta_{\rm d}^{\rm low}[M_{\odot}]$ / $\delta_{\rm d}^{\rm high}[M_{\odot}]$ & Smooth scale of the lower / upper side of dip & $LU(10^{-2},2)$  \\
$\{f_j\}_{j=2}^{14}$ & Interpolation values of perturbation function & $\mathcal{N}(0,1)$ \\
constraints & & $m_{\rm d,1}<m_{\rm d,h}$ \\
\hline
$R_0[{\rm Gpc}^{-3}~{\rm yr}^{-1}]$ & Local merger rate  density & $U(0,100)$ \\
$\gamma$ / $\kappa$ & slope of the power-law regime for the rate evolution before / after $z_{\rm p}$ & $U(0,12)$ \\
$z_{\rm p}$ &redshift turning point between the power-law regimes with $\gamma$ and $\kappa$ & $U(0,4)$\\
\hline
$H_0[{\rm km~s^{-1}~Mpc^{-1}}]$  &Hubble constant &  $U(10,200)$ \\
 $\Omega_{\rm m}$ & present-day dimensionless matter densities  & 0.3065 \\
 \hline
\multicolumn{3}{c}{\bf For BBH models}\\
$m_{\rm min}[M_{\odot}]$   & The minimum   mass & $U(2,10)$  \\
$\alpha$  & Slope index of the power-law & $U(1.5,12)$ \\
$\delta_{\rm m}[M_{\odot}]$ & Smooth scale of the mass lower edge & $U(10^{-2},10)$  \\
\hline
\multicolumn{3}{c}{\bf For Gaussian peaks (Fullpop-4.0, MLTP, MLTP paired)}\\
$\mu_{\rm g}^{\rm low}[M_{\odot}]$/$\mu_{\rm g}^{\rm high}[M_{\odot}]$   & Location of the first / second peak & $U(5,15)$ / $U(15,45)$  \\
$\sigma_{\rm g}^{\rm low}[M_{\odot}]$/$\sigma_{\rm g}^{\rm high}[M_{\odot}]$   & Width of the first / second peak & $U(5,15)$ / $U(15,45)$  \\
$r_{\rm g}$ &  Fraction of sources in all the peaks & $U(0,1)$ \\
$r_{\rm g}^{\rm low}$ &  Fraction of sources in the first peak & $U(0,1)$ \\
\hline
\hline
\end{tabular}
\\
\begin{tabular}{l}
Note: $U$, $LU$,  $\mathcal{N}$, and $\mathcal{G}$ are for Uniform, LogUniform, Normal distribution, and Gaussian distribution.
\end{tabular}
\end{table*} 

\section{Additional Restuls}\label{app:results}
\begin{figure}[ht!]
\centering
  \includegraphics[width=0.48\textwidth]{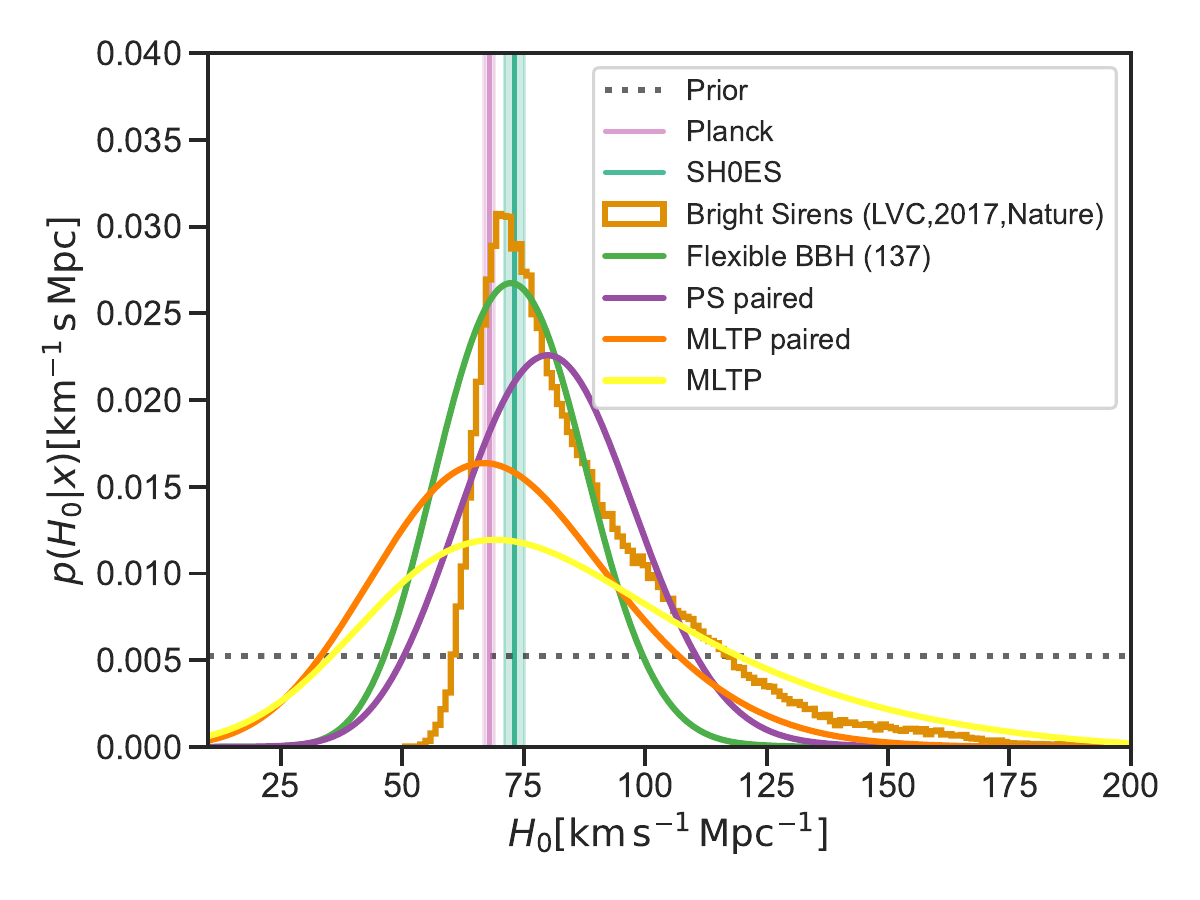}
\caption{Hubble constant posteriors inferred from 137 BBH events using spectral sirens with four models (Flexible-BBH, PS paired, MLTP paired, and MLTP), together with the constraint from the bright siren GW170817 \citep{2017Natur.551...85A}. Vertical lines indicate the reference values from Planck and SH0ES \cite{2020A&A...641A...6P,2022ApJ...934L...7R}, highlighting the Hubble tension.}
     \label{Fig:BBH_H0}
\end{figure}

\begin{figure}[ht!]
\centering
  \includegraphics[width=0.5\textwidth]{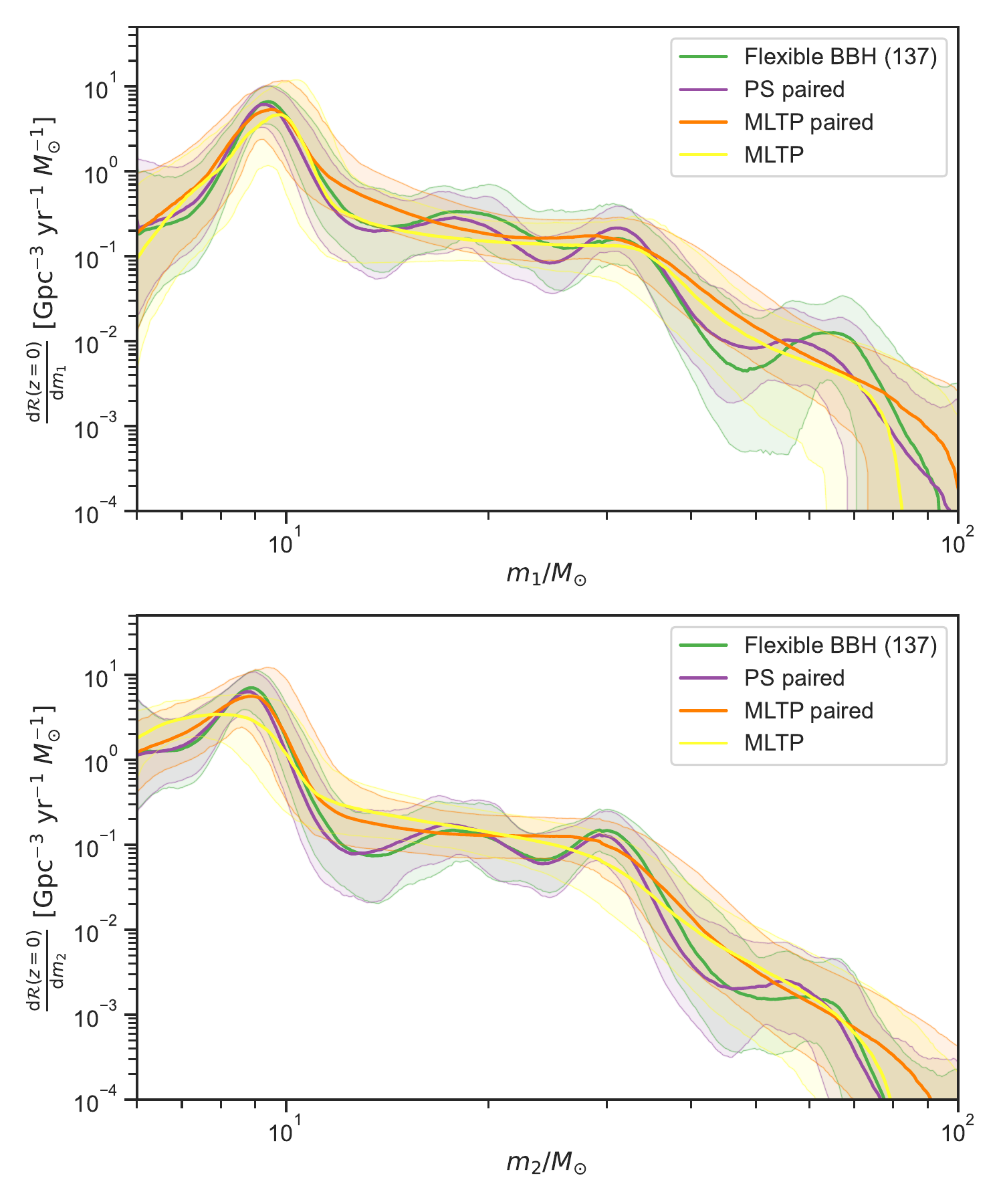}
\caption{Inferred mass distributions of primary and secondary BHs under the Flexible-BBH, PS paired, MLTP paired, and MLTP models. Solid lines and shaded areas represent median values and 90\% credible intervals, respectively. The MLTP model fails to capture the peaks at $\sim 10\,M_\odot$ and $\sim 30\,M_\odot$ in the secondary-mass distribution.}
     \label{Fig:BBH_mass}
\end{figure}

\begin{figure}[ht!]
\centering
  \includegraphics[width=0.48\textwidth]{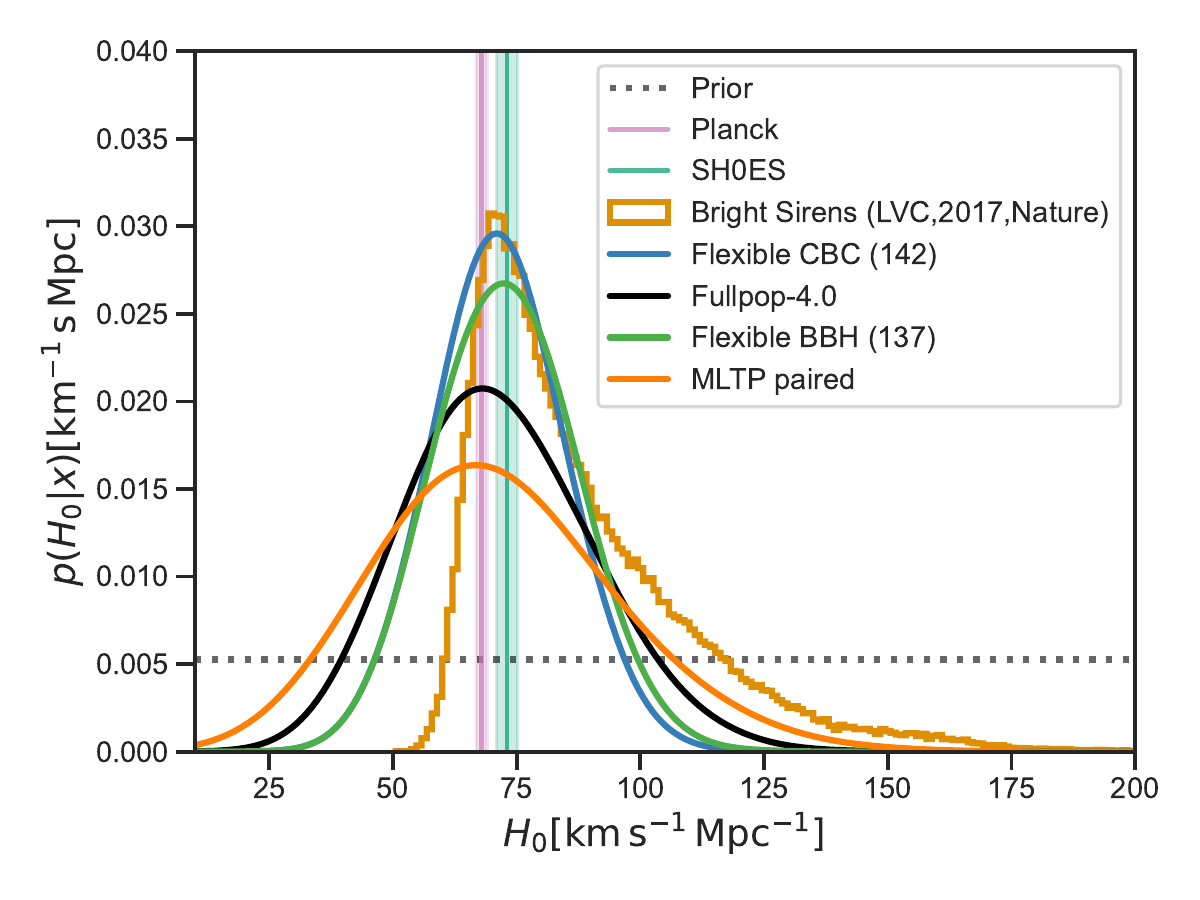}
\caption{Posterior distributions of the Hubble constant inferred from spectral sirens using the Flexible-CBC, Flexible-BBH, MLTP paired, and MLTP models, along with the constraint from the bright siren GW170817 \citep{2017Natur.551...85A}. Vertical lines mark the reference values from Planck and SH0ES~\cite{2020A&A...641A...6P,2022ApJ...934L...7R}, highlighting the Hubble tension.}
     \label{Fig:CBC_BBH_H0}
\end{figure}

\begin{figure*}[ht!]
\centering
  \includegraphics[width=0.9\textwidth]{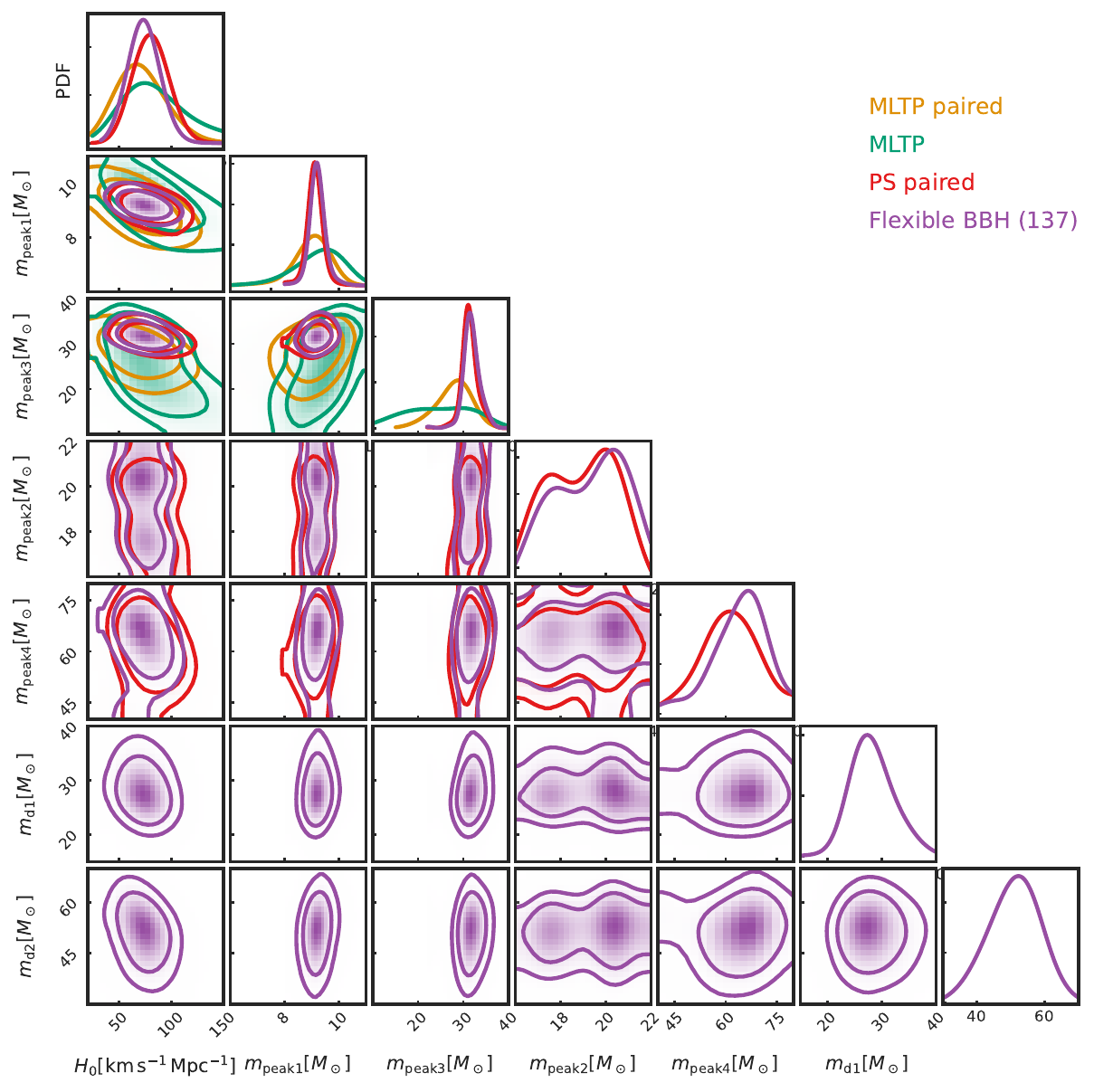}
\caption{Posterior distributions of the Hubble constant and the parameters describing the BBH mass distribution, for the Flexible-BBH, PS paired, MLTP paired, and MLTP models. The four peaks correspond to the maxima of the cubic spline functions for the Flexible-BBH and PS paired models. Among them, peak1 and peak3 represent the central values of the two Gaussian components ($\mu_{\rm g}^{\rm low}$ and $\mu_{\rm g}^{\rm high}$) as defined in Fullpop-4.0~\cite{2025arXiv250904348T}.}
     \label{Fig:BBH_corner}
\end{figure*}

\begin{figure*}[ht!]
\centering
  \includegraphics[width=0.45\textwidth]{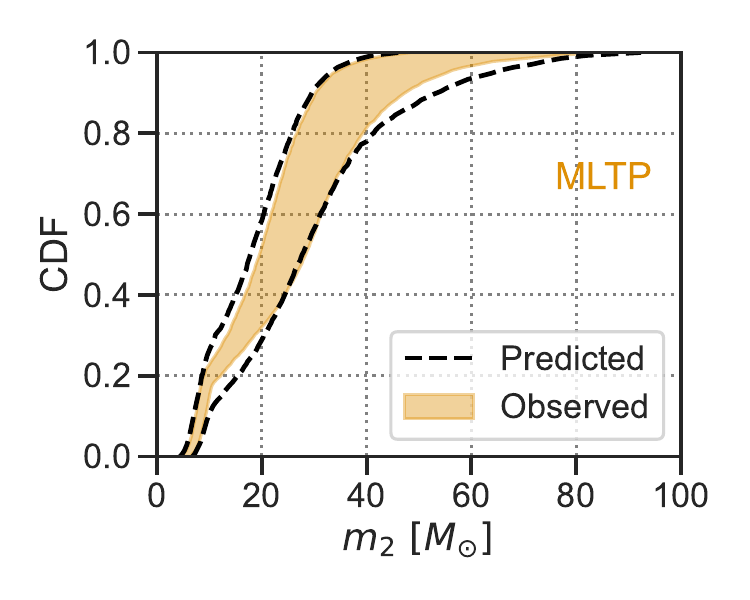}
  \includegraphics[width=0.45\textwidth]{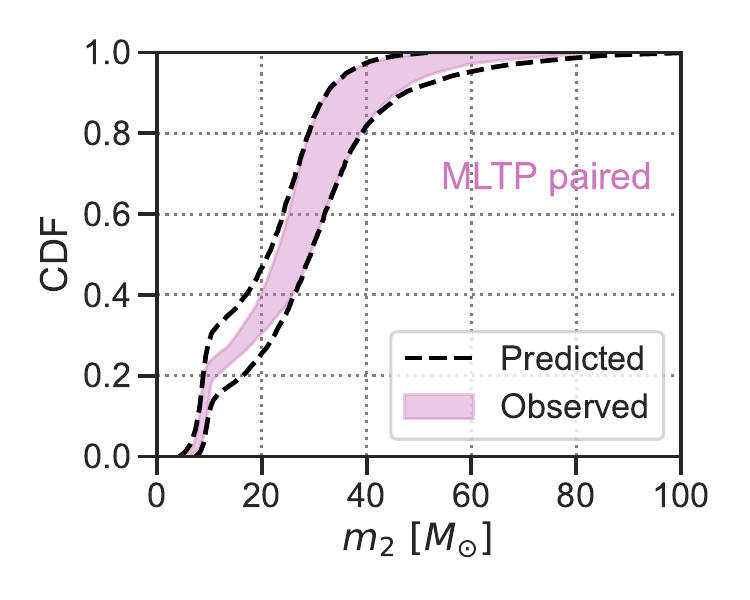}
  \includegraphics[width=0.45\textwidth]{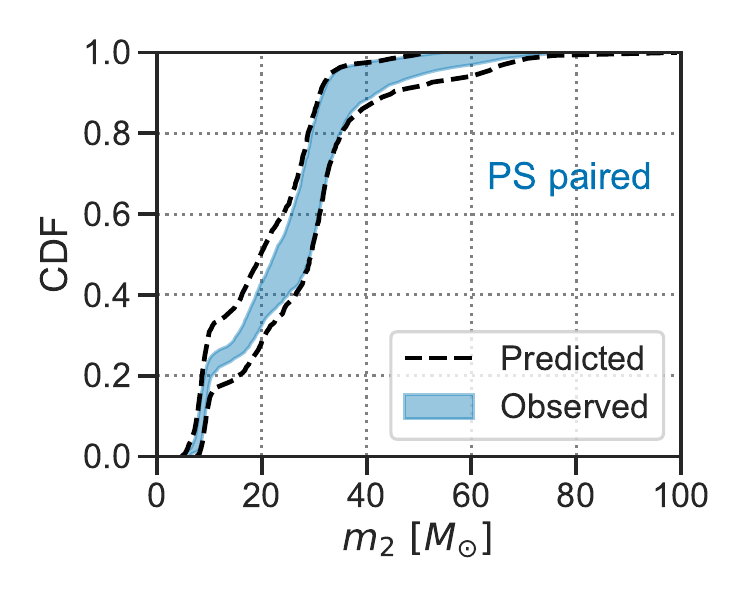}
  \includegraphics[width=0.45\textwidth]{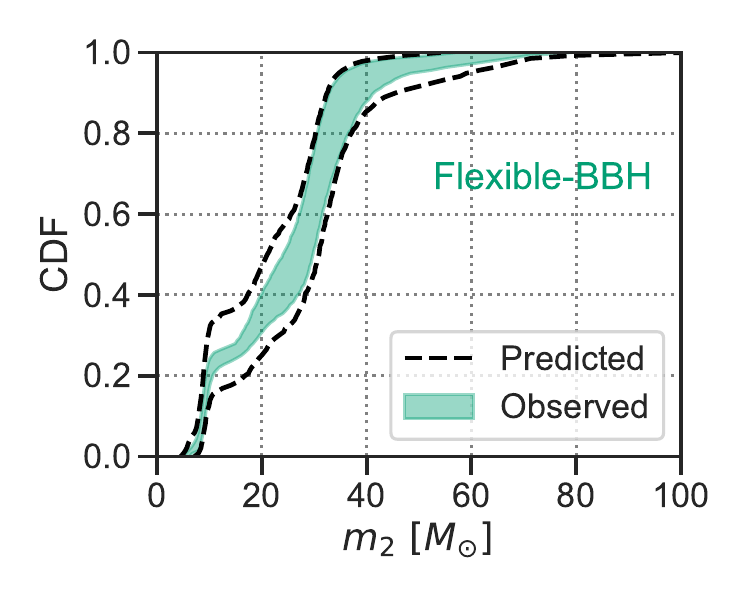}
\caption{Posterior predictive check: cumulative distribution functions (CDFs) of the observed secondary mass distribution for the MLTP, MLTP paired, PS paired, and Flexible-BBH models. Shaded regions (dashed lines) represent the observed (predicted) event distributions. All bands indicate 90\% credible intervals. The observed band from the MLTP model lies slightly outside the predicted region, indicating its weakness relative to the other models. The Flexible-BBH, PS paired, and MLTP paired models provide a better fit than the truncated model, as their shaded regions overlap entirely with the dashed bands, owing to their greater flexibility in capturing additional features in the secondary-mass distribution.}
     \label{Fig:PPC}
\end{figure*}

%\subsection{Impact of the secondary-mass function}\label{app:secondary}
LVKC~\cite{2025arXiv250904348T} suggest that adopting a multi-population mass model such as Fullpop-4.0 significantly improves the spectral siren constraints on $H_0$, even with the inclusion of just five additional candidates containing at least a potential neutron star. They find that the Fullpop-4.0 model yields an improvement of $\sim51\%$ over the MLTP model. Using the dedicated BBH mass function described in Section~\ref{app:BBH_model}, we show that this improvement arises not only from the inclusion of the five low-mass events, but also from a more accurate modeling of the pairing function and, consequently, the secondary-mass function. For example, if we apply a pairing function to the MLTP mass function (i.e., the MLTP paired model), we obtain $H_0 = 70^{+27}_{-24}~\text{km}\,\text{s}^{-1}\,\text{Mpc}^{-1}$, which corresponds to a 30.2\% improvement over the original MLTP model. Meanwhile, the improvement of Fullpop-4.0 relative to the MLTP paired model is 21.5\%, which stems from the inclusion of the five low-mass events. Note that both the Fullpop-4.0 and MLTP paired models incorporate a pairing function, whereas the original MLTP model has a secondary-mass function conditioned on $m_1$.

Figure~\ref{Fig:BBH_mass} shows the inferred mass distributions of primary and secondary BHs. The secondary mass distribution exhibits distinct features between the MLTP and MLTP paired models: the MLTP model may have neglected potential peaks at $\sim 10\,M_\odot$ and $\sim 30\,M_\odot$ in the secondary-mass function, which leads to a looser measurement of $H_0$ compared to the MLTP paired model. The posterior predictive check is presented in Figure~\ref{Fig:PPC}, which shows that the observed band from the MLTP model lies slightly outside the predicted region, indicating the weakness of the MLTP model relative to the other models.

\clearpage
\bibliography{references}{}
\bibliographystyle{aasjournalv7}



\end{CJK*}
\end{document}